\documentclass[%
 reprint,
 amsmath,amssymb,
 aps,
floatfix,
]{revtex4-1}

\usepackage{graphicx}% Include figure files
\usepackage{dcolumn}% Align table columns on decimal point
\usepackage{bm}% bold math
\usepackage{epstopdf}
\usepackage{comment}
\usepackage{float}
\usepackage{MnSymbol}

\begin{document}

\preprint{APS/123-QED}

\title{On the abnormal temperature dependent elastic properties of fused silica irradiated by ultrafast lasers}% Force line breaks with \\

\author{Pieter Vlugter}
 \email{pieter.vlugter@epfl.ch}
\author{Yves Bellouard}
\affiliation{Galatea Lab, STI/IEM, Ecole Polytechnique F\'{e}d\'{e}rale de Lausanne (EPFL), 2002 Neuch\^{a}tel, Switzerland}%Lines break automatically or can be forced with \\

\date{\today}% It is always \today, today,

\begin{abstract}
\noindent Materials with thermal-invariant elastic properties of materials are of interest for resonant device frequency and dimensional stability of precision devices. Here, we demonstrate that the temperature coefficient of elasticity (TCE) of amorphous silica  can be locally reduced using femtosecond laser irradiation. Notably, a self-organized laser-induced modification shows a decrease of the TCE beyond 50\%. The origin of this dramatic decrease is attributed to the presence of sub-micron permanent densified zones within the fine multi-layers structure of the laser-modified zone.
\end{abstract}

\pacs{Valid PACS appear here}% PACS, the Physics and Astronomy
\maketitle
%general about beta

\begin{figure}[h!]
\centering
\includegraphics[width=0.8\linewidth]{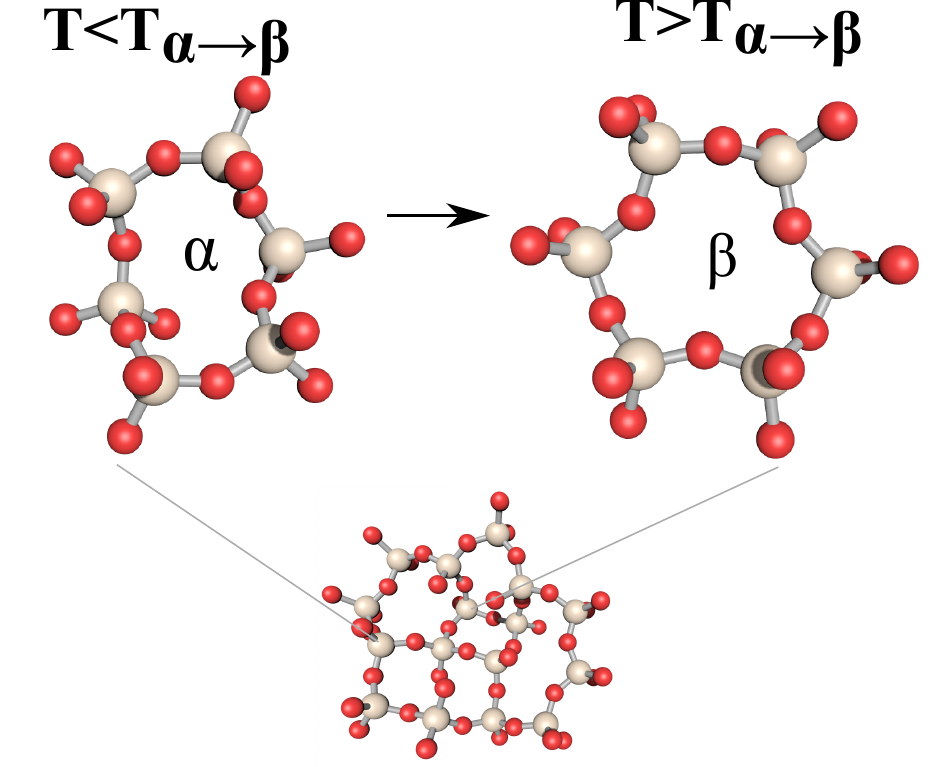}
\caption{An impression of the six-membered ring's conformation changes into an $\alpha$ to the $\beta$ form due to temperature increase, adapted from \cite{Huang2004-qi,Deng2020-jo}. Where the red and white atom represents silicon and oxygen, respectively.}
\label{Fig_DubblePotentialWell}
\end{figure}

\noindent Amorphous silica a-SiO$_2$ has unique behavior among materials in general. Unlike many other glasses and the vast majority of materials in general, it exhibits a positive temperature coefficient of elasticity (TCE), i.e. its Young's modulus increases with increasing temperature as reported by R.B. Sosman in 1927 \cite{Uhlmann2012-xx}. The cause for this unusual behavior has since then been disputed. One recurring idea, proposed by Babcock in 1954 \cite{Babcock1954-kg}, is that a-SiO$_2$ contains two types of structure, coexisting within the glass, each of which behaves normally locally, where their relative amounts change continuously with temperature and pressure. According this model, the material undergoes a reversible displacive  transformation from an elastic weak state ($\alpha$-form) to a stiffer one ($\beta$-form), similar to cristobalite, one of the cristalline polymorphs of SiO$_2$. While for the crystalline phases, the transformation, $\alpha \rightleftharpoons \beta$, occurs in a relative small temperature window, it would transform over a broad temperature range in the case of a-SiO$_2$. It is widely acknowledged that the structure of a-SiO$_2$ is not fully random and possess a short-range order (< 0.5nm), while the medium-range structure (0.5 nm-1.5 nm) is consisting in a network of rings of various sizes formed by SiO$_2$ tetrahedra   \cite{Zachariasen1932-ou,King1967-mp,Galeener1982-uj}, where the six-membered ring structures is the most abundant member. According to molecular dynamics simulations done by L. Huang \emph{et al.}, these six-number ring-structures are specifically responsible for the stiffening effect, by conformation changes of the ring structures \cite{Huang2004-qi}, see Figure \ref{Fig_DubblePotentialWell}. The four coordinated tetrahedra SiO$_2$ act as a rigid body unit \cite{Hammonds1996-jh}. During the short-range thermal expansion, the distance between Si and O atom of the four coordinated tetrahedra SiO$_2$ molecule \cite{Shi2020-ov}, causes the six-membered ring structures to transform from an asymmetric ($\alpha$-form) to a more symmetric, and stiffer, configuration ($\beta$-form).

% Mention maybe first that it is via an external stimulus 
a-SiO$_2$'s TCE is altered while undergoing irreversible transformations towards different metastable amorphous states, a transformation mechanism coined by M. Grimsditch as ’amorphous polyamorphism’ \cite{Grimsditch1984-yk}. Two amorphous state families are primarily identified, the low (LDA) and high density amorphous states (HDA), respectively. Interestingly, like other abnormal properties (the boson peak \cite{Nakayama2002-hy}, softening upon compression \cite{Bridgman1925-ln}), permanently densified a-SiO$_2$ reduces the magnitude of the TCE. Furthermore, unlike amorphous ice, a-SiO$_2$ possesses a broad range of states, within the categories of LDA and HDA states, that can be reached through various thermodynamic paths \cite{McMillan2007-jy} or using various intense irradiation methods that stimulate bond-breaking processes and localized reordering \cite{Schenker1997-hs,Reghioua2019-kz}. As there is no unique path towards a given state, different densification methods, such as fast neutron exposure \cite{Strakna1963-tu}, thermal quenching \cite{Le_Parc2006-oz} and cold- and hot-compression \cite{Guerette2018-fs}, will inherently lead to distinct densification state, amorphous structure and TCE reduction level. 

Here, we focus our attention on the local changes of a-SiO$_2$' TCE resulting from femtosecond laser pulses  bulk-exposure. Thanks to their high peak-power that triggers non-linear absorption processes, and despite having a wavelength for which the material is transparent, these lasers are capable of modifying locally transparent materials only in the focal volume \cite{Stuart1995-iq,Glezer1997-kq}. In a-SiO$_2$, the modified volume is associated with microscale material property changes, e.g. refractive index \cite{Davis1996-ae}, chemical susceptibility \cite{Marcinkevi_Ius2001-dm}, thermal conductivity \cite{Bellouard2008-uz}, thermal expansion coefficient \cite{Vlugter2019-ye} and Young's modulus \cite{Bellouard2006-ad, Vlugter2020-xl}. These laser-induced physical properties changes are particularly useful in the context of monolithic micro-devices and led to the creation of monolithic devices, e.g. complex integrated photonics circuits \cite{Minoshima2002-in}, tuned resonators \cite{Yang2017-ns, Vlugter2020-xl},  optomechanical devices \cite{Bellouard2005-uv} and optofluidics \cite{He2014-jj,Haque2014-hv,Schaap2011-kl}, made from a single piece of material and functionalized through laser exposure sequences were reported. Specifically, a-SiO$_2$ is attractive for its remarkable material properties, such as its low thermal expansion, its low optical losses over a broad spectral range as well as its low mechanical losses and high thermal stability. Unfortunately, its TCE remains not negligible, $\sim$210 ppm \cite{Spinner1956-bb,Bao2016-sp}, especially in contrast with nickel-based alloys commonly referred as Elinvar or Nivarox - that have a TCE around and below ppm levels \cite{Kikuchi1987-qb}. 
To further enhance a-SiO$_2$ applicability, we explore how femtosecond laser-irradiated zones can also be used to lower the intrinsic TCE value of the material. Specifically, we investigate two types of laser-induced modifications: one, commonly referred as 'type I', consisting of an homogeneous modification throughout the exposed zone, and a second one, usually called 'type II',  consisting of self-organized nanoplanes parallel one to another.

% Experimental details
\section*{Experimental methods}
\begin{figure*}[b]
\centering
\includegraphics[width=\linewidth]{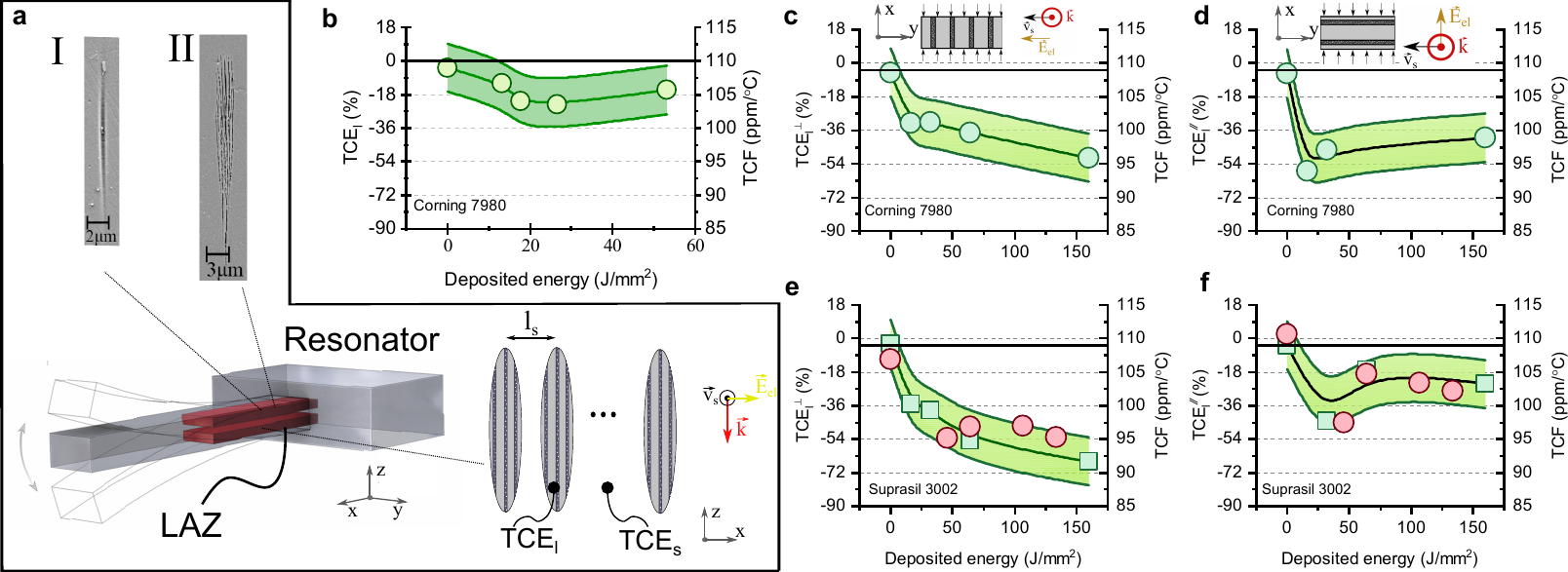}
\caption{\textbf{a}) a schematic of the mechanical resonator with in red the laser affected zones (LAZ). The insets above show two scanning electron microscope (SEM) images, typical cross-section of type I and II modifications. A schematic further depicts a set of nanograting modifications spaced by $l_s$, and consisting of alternating porous (TCE$_{pl}$) and non-porous layers (TCE$_{nl}$) planes. $\vec{v}_s$ is the laser writing direction, $\vec{k}$ is the optical propagation axis and $\vec{E}_{el}$ is the direction of the electric field defining the polarization of the laser. The measured TCF and related TCE are shown in graphs \textbf{b}-\textbf{f}). \textbf{b} refers to the case of a type I modification.  \textbf{c} and \textbf{e}, show the result in the case of a type II modification for nanogratings oriented along the cantilever long-axis while, \textbf{d} and \textbf{f}) for a type II modification, but here with the nanogratings oriented in the transverse direction to the cantilever long-axis.}
\label{Fig_Mainresults}
\end{figure*}

\noindent TCE properties are retrieved using a method based on measuring the mechanical resonance of laser modified cantilevers as a function of temperature. We take advantage of a process based on femtosecond laser exposure and chemical etching to fabricate  micro-cantilevers. The volume inside those cantilevers are exposed to ultrafast laser as depicted in Figure \ref{Fig_Mainresults}a. The fabrication of these structures, dimensions and exposure strategy are further detailed in \cite{Vlugter2020-xl}. 

The resonance frequency dependency on temperature, also referred as temperature coefficient of frequency TCF, expressed as function of material properties \cite{Melamud2009-bb} is;
\begin{eqnarray}
    \textrm{TCF}=\frac{1}{f_0}\frac{\partial f}{\partial T}&=&(\textrm{CTE}+\textrm{TCE})/2 \label{equa_TCF}\\
                &=& \left(\textrm{CTE}+\frac{1}{E_0}\frac{\partial E}{\partial T}\right)/2 \nonumber
\end{eqnarray}
Where $f$ is the resonance frequency and $E$ is the Young's modulus - the subscript 0 refers to measured property at room temperature. Additionally, the TCF depends on the thermal expansion coefficient (CTE). Nevertheless, the low CTE of a-SiO$_2$ ($\sim$ 0.5 ppm/$^\circ$C) makes this dependence negligible compared to the TCE ($\sim$ 220 ppm/$^\circ$C) which is three orders of magnitude higher \cite{Spinner1956-bb,Bao2016-sp}. As we will see later, the laser induced CTE changes, reported in \cite{Vlugter2019-ye}, are likewise negligible compared to the laser induced TCE changes. 

\onecolumngrid\

\twocolumngrid\

\clearpage
\noindent The resonator volume is not entirely and not homogeneously exposed to the laser. Instead, the modified zones consists of a set of lines, spaced by a distance $l_s$, and located just below the surface (Figure \ref{Fig_Mainresults}a). The TCE extracted from the measured TCF expressed in equation \ref{equa_TCF} is the sum of pristine (TCE$_s$) and laser affected zones (TCE$_l$). The change of TCE in the modified volume TCE$_l$ is estimated based on a mathematical model, detailed in the Appendix A.

The temperature is controlled and cycled between 10 $^\circ$C to 40 $^\circ$C, using a dedicated environmental chamber described in \cite{Vlugter2019-ye}. The cantilever is mounted on a piezoelectric-actuator that excites it on its resonant frequency. The resonator's response is measured using a triangulation sensor (Keyence LKH022) and a phase-lock-loop scheme is applied to track the resonance frequency changes during temperature cycles. Type I modifications are produced using a femtosecond laser system emitting 50 fs-pulses at 850 nm (Satsuma seeding an OPA from Amplitude), while type II modifications are obtained with an Yb-doped fiber-amplified system emitting 380 fs-pulses  at 1030 nm (Yuzu from Amplitude). In the sequel, the pulse energies used to obtain the modified zone are kept constant (250 nJ for type I and 240 nJ for type II), while the exposure dose is varied by changing the scanning speed. 
As for the substrates, we consider two types of a-SiO$_2$: one with a high OH content (Corning, ref. 7980) and one with a low OH content (Heraeus, Suprasil 3002).

\section*{Results}
\noindent Figure \ref{Fig_Mainresults}b shows the TCF and the corresponding TCE$_l$ variation as a function of the deposited energy for type I modifications. With these experimental conditions, we observe a TCF decrease of 7 ppm/$^\circ$C and estimate a TCE decrease of about 20\%. 
\noindent Figure \ref{Fig_Mainresults}c and d, considers  the TCF and estimated TCE versus the deposited energy are depicted for a type II modifications for two opposite nanograting orientations. The change of TCE shows a distinct behavior in both cases. A saturating behavior for the case displayed in Figure \ref{Fig_Mainresults}d and f, corresponding to a transverse nanograting orientation with respect to the cantilever long-axis is found, while a linear decay for the case in Figure \ref{Fig_Mainresults}c and e is observed and corresponds to nanogratings aligned along the cantilever long-axis. This observation is useful to understand better what the origin of the reduction is and will be discussed more in depth in the following section. 
\noindent For type II modifications, the TCE reductions are of larger magnitude compared to type I modifications. For nanograting aligned along the cantilever long-axis (Figure \ref{Fig_Mainresults}c and e)  the reduction is eventually reaching values $\Delta$TCE$\sim$60$\%$ for a deposited energy of 160 J/mm$^2$. 
The LAZ's CTE changes are negligible compared with the TCE changes, based on data reported in \cite{Vlugter2019-ye}, which confirms our earlier stated hypothesis. 

\section*{Hypothesis towards the origin of the TCE reduction}
\begin{figure}[h!]
\centering
\includegraphics[width=1\linewidth]{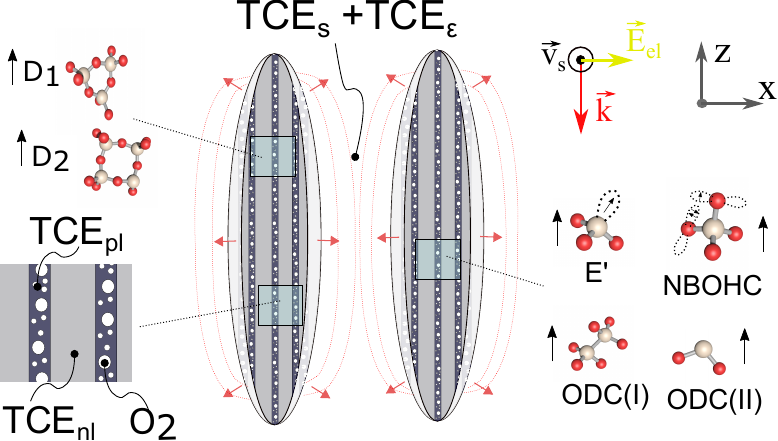}
\caption{Cross-section of two adjacent laser-written tracks with different regions, characteristics of a type II modification (so called 'nanograting' structure). TCE$_{nl}$ and TCE$_{pl}$ are the temperature coefficient of elasticity of the non-porous and porous layers, respectively.  TCE$_{s}$ and TCE$_\epsilon$ are the temperature-dependence coefficient of elasticity of the pristine material and the strained region, respectively. The red contour map illustrates the strain induced by the laser induced volume expansion \cite{Champion2013-nn}. Further, the increase of point defect density is illustrated and the increase of the three- and four-membered rings.}
\label{Different_TCE}
\end{figure}

\noindent Considering the nanogratings morphology \cite{Lancry2013-fv} and the laser patterns considered here, several features could be responsible for the observed TCE reduction. Namely, the porous layers (TCE$_{pl}$) and/or the non-porous layers (TCE$_{nl}$), permanent densifications, elastically deformed material surrounding the LAZ (TCE$_{\epsilon}$) and laser induced point defects are possible candidates to explain the observed behavior. Figure \ref{Different_TCE}a illustrates the characteristic laser induced material changes and are the possible origins for the TCE reduction. 

\subsection*{Elastic deformation}
\noindent The porous character is responsible for a net volume expansion of the modified volume \cite{Champion2012-ny}. Consequently, the surrounding unexposed material is elastically densified. From compression studies in the elastic regime, it is known that the athermal Young's modulus softens \cite{Deschamps2014-jt, Bridgman1925-ln}. This could indicate that the symmetric ring structures undergo a displacive transformation towards a weaker asymmetric form, due to an external pressure \cite{Huang2004-qi}. Following this hypothesis, one can expect that this permanent external pressure prevents this weaker asymmetric form to transform back into its stiffer symmetric configuration. As a consequence, the TCE would reduce as the stiffening mechanism (described in the introduction section) is obstructed due to the presence of post-laser exposure residual permanent stress. To support this hypothesis, studies on other abnormal material properties of SiO$_2$ (i.e. boson peak \cite{Deschamps2011-gy, Deschamps2009-bw,Hemley1986-pm}) show that elastic densification does reduce the magnitude of these abnormal behaviors. Unfortunately, to the best of our knowledge, no experimental result has been reported related to the correlation between elastic deformation and TCE. 

\begin{figure}[h!]
\centering
\includegraphics[width=1\linewidth]{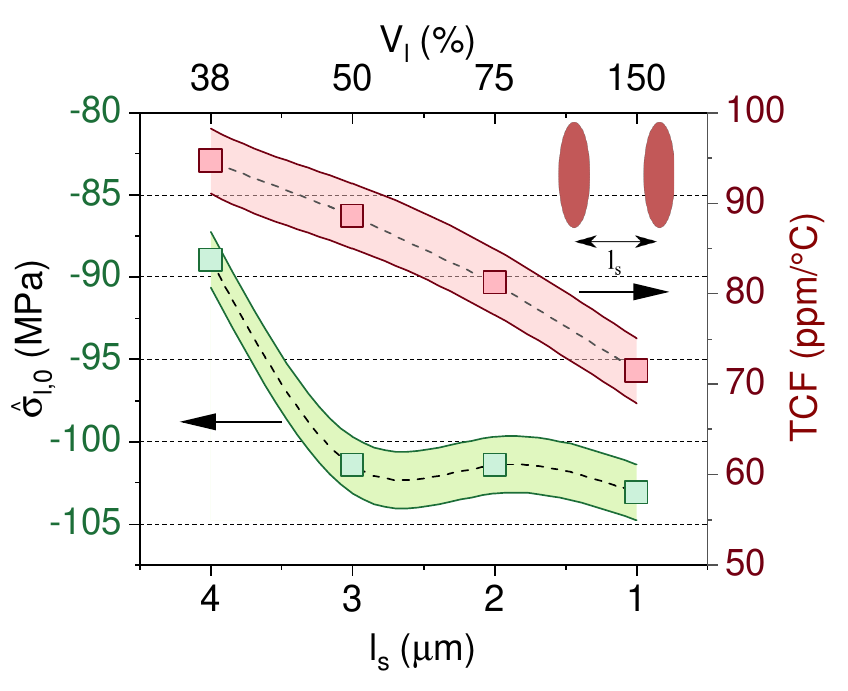}
\caption{Stress ($\hat{\sigma}_{l,0}$) and TCF versus laser-written line-spacing.The exposure conditions are; pulse energy 300 nJ, repetition rate 750 kHz, scanning velocity 5 mm/s, pulse width 300 fs and laser-wavelength 1030 nm.}
\label{Fig_linespacingvsTCFandStress}
\end{figure}

In the sequel, we make a comparative study between an experiment for probing the stress present in LAZs and the one to measure the TCE as described earlier. To probe the stress, a bimorph deflection experiment is used, like in \cite{Champion2012-ny}, where the tip displacement of an laser-exposed cantilever is related to the stress using a modified Stoney's equation as detailed in Appendix B. 

Unlike the previously discussed experiments, the exposure conditions (i.e. pulse energy and exposure dose) are kept constant, but the line spacing between individual laser-written lines is varied ($l_s$ in Figure \ref{Fig_Mainresults}a). This parameter effectively controls the average stress present in laser-written patterns, see Figure \ref{Fig_linespacingvsTCFandStress}. On the one hand, the stress seems to saturate for line-spacing below 4 $\mu$m (green curve in Figure \ref{Fig_linespacingvsTCFandStress}), which is in agreement with our  previous observations \cite{Vlugter2019-ye}. On the other hand, the TCF does not follow the same trend, but rather, seems to be linearly proportional with line-spacing. There is therefore no obvious correlation between the elastic deformation of the unexposed material in between laser-written line and the reduction of TCE. This agrees well with the MD-simulations reported by Huang \emph{et al.} \cite{Huang2004-qi}, where they concluded that the $\alpha$-$\beta$ transitions are saturated beyond the pressure window ranging from -9 to 6 GPa. Therefore, for the pressure levels obtained in our experiments, a sufficient amount of asymmetric ring-structures can be thermally transformed to his stiffer counterpart.  

\subsection*{Porous and non-porous layer}
\noindent At this point of the discussion, it is excluded that the elastic deformation has an effect on the observed TCE reduction. To quantify whether porous or non-porous layers are contributing to the reduction of the TCE, we revisit the results regarding the two orthogonal nanograting orientations, depicted in Figure \ref{Fig_Mainresults}. For the nanogratings aligned along the loading direction (Figure \ref{Fig_Mainresults}, the TCE of the non-porous layers (TCE$_{nl}$) are dominant, since it is equivalent to a mechanical parallel configuration. Furthermore, the non-porous layer is nearly a factor ten stiffer than porous layer and occupies nearly 90\% of the volume \cite{Vlugter2020-xl}. For the orthogonal arrangements (Figure \ref{Fig_Mainresults}d and f), the porous layers contribution is this time prominent, as this arrangement of layers is equivalent to in-series connected stiffness. This reasoning is supported by estimates made in equation \ref{equa_BetaPol0} that considers nanogratings aligned along the loading direction and equation \ref{equa_BetaPol90}, for which they are aligned transversely. For $V_{nl}\sim$ 0.9,$V_{pl}\sim$ 0.1 (\cite{Bricchi2006-qy}) and $E_{pl}\sim$ $E_{nl}$0.1 (\cite{Vlugter2020-xl}), equation \ref{equa_BetaPol0} reveals that the behavior of $\textrm{TCE}^\perp_l$ in Figure \ref{Fig_Mainresults}c and e is mostly governed by non-porous layers contributions to the TCE, while for an orthogonal nanograting orientation, equation \ref{equa_BetaPol90}, both non- and porous- layers contribute nearly equally, resulting in the behavior observed in Figure \ref{Fig_Mainresults}d and f.

\begin{align}
\textrm{TCE}^\perp_l=&\frac{\textrm{TCE}_{nl}V_{nl}E_{nl}+\textrm{TCE}_{pl}V_{pl}E_{pl}}{E_{nl}V_{nl}+E_{pl}V_{pl}} \hspace{0.5cm} \rightarrow \nonumber \\
\label{equa_BetaPol0}\\
\textrm{TCE}^\perp_l \sim& \textrm{TCE}_{nl}0.9+\textrm{TCE}_{pl}0.01 \nonumber
\end{align}

\begin{align}
\textrm{TCE}^\Neswline_l=& \frac{\textrm{TCE}_{nl} V_{nl} E_{pl}+\textrm{TCE}_{pl} V_{pl} E_{nl}}{V_{nl} E_{pl}+V_{pl} E_{nl}} \hspace{0.5cm} \rightarrow \nonumber \\
 \label{equa_BetaPol90}\\
 \hspace{0.5cm} \textrm{TCE}^\Neswline_l\sim& \textrm{TCE}_{nl}0.9+\textrm{TCE}_{pl}\label{equa_BetaPol90} \nonumber 
\end{align}

\subsection*{Annealing}
\noindent To further test the hypothesis that permanently densified zones (i.e. zones within the laser modified volume and in between nano-porous layers) are the main source for TCE reduction, we investigated the behavior of the cantilever after various thermal annealing steps. Depending on the temperature it is performed, thermal annealing of laser affected zones have different effects. For instance,  for low annealing temperatures ($T_{ann}<\textrm{500}$ $^\circ$C), laser induced point defects (e.g., dangling bonds, non-bridging oxygen hole centers (NBOHC), etc.) are specifically altered \cite{Skuja2005-rk,Sun2000-jm}, while for intermediate temperatures (500 $^\circ$C $<T_{ann}<\textrm{900}$ $^\circ$C), permanent densification is quenched \cite{Guerette2018-fs, Cornet2019-cr}. Finally, the porous layers are removed for annealing temperatures beyond the so called "glass annealing temperature" \cite{Cavillon2020-lb}, which for the Corning 7980 glass is found at $T_{ann}\sim$1042 $^\circ$C. Therefore, we logically expect changes in the TCE depending on the annealing temperature that may provide further hints about the origin of the TCE reduction.

\begin{figure*}[h!]
\centering
\includegraphics[width=\linewidth]{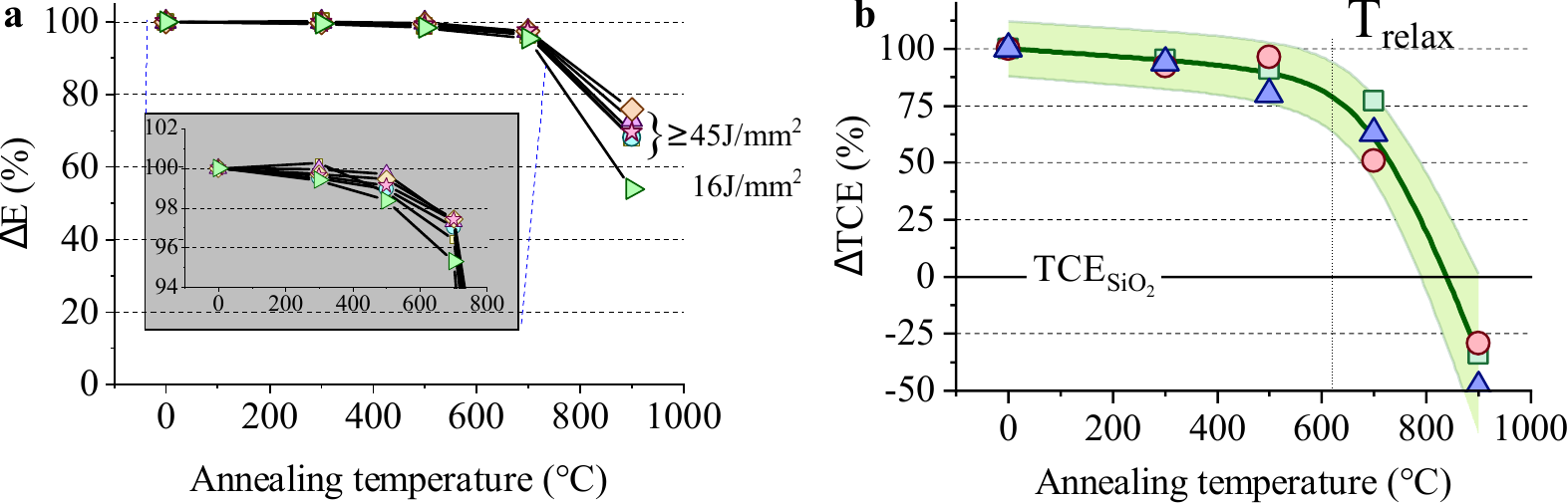}
\caption{\textbf{a}) The change in athermal Young's modulus ($\Delta$E) versus the annealing temperature. \textbf{b}) The  change in LAZ’s temperature-elastic coefficient ($\Delta TCE$) versus annealing temperatures.}
\label{Fig_TCE_AND_EvsTann}
\end{figure*}

In this set of experiments, we characterized the athermal Young's modulus and the TCE of post-annealed resonators under various conditions. In practice, the resonators are placed in a nitrogen-purged furnace and the temperature is elevated to a set of characteristics levels $T_{ann}$. Heating and cooling rates are set below 1 $^\circ$C/min, to avoid additional residual stress and undesired transient effects. To study the fine changes of the TCE as function of the annealing temperature, we reuse the specimens with close-spaced exposure patterns discussed in Figure \ref{Fig_linespacingvsTCFandStress}. For investigating the changes in athermal Young's moduli, we reuse the samples discussed in Figure \ref{Fig_Mainresults}d. The athermal modulus is known to decay as function of the deposited in an inversely proportional manner  \cite{Vlugter2020-xl}. This reduction of the elastic properties is attributed to the porous planes  \cite{Vlugter2020-xl}. 
 Figure \ref{Fig_TCE_AND_EvsTann}a and b depicts the athermal Young's modulus (E) and TCE normalized relative changes, respectively. The references for normalization are the non-annealed laser-exposed material, meaning that 100\% corresponds to the non-annealed case and 0\% to the unexposed one. 
 
 Overall, the laser-induced TCE and Young modulus changes remain almost unaffected ($<$2\%) for increasing annealing $T_{ann}$ up to 600 $^{\circ}$C. This indicates that point defects like E' centers, NBOHC, ODC(I) and ODC(II) do not have a noticeable effect on the reduction of the TCE as their concentration  gradually deplete at these annealing temperatures and durations \cite{Sun2000-jm,Zoubir2006-kg,Witcher2013-fl}. Likewise, interstitial oxygen, found in the Raman spectrum \cite{Lancry2013-fv,Bellouard2016-ck}, disappears at these annealing temperatures \cite{Zimmermann2017-an}. In \cite{Sun2000-jm}, it is suggested that interstitial oxygen and E' centers interact and form peroxy radicals (for $T_{ann}<$400 $^{\circ}$C). Although these chemical rearrangements highly likely have an effect on the thermal elastic behavior, its significance is too small to measure due to a relative low concentration.          

The annealing experiments to $\Delta$E (Figure \ref{Fig_TCE_AND_EvsTann}a) indicate that porous planes survive at least up to a temperature of $T_{ann}<$900 $^{\circ}$C. This high thermal-stability of the porous planes have been also observed elsewhere using a different methodology \cite{Zhang2014-kt,Wei2020-ut} and is further extensively studied and modeled in \cite{Cavillon2020-lb,Wang2021-ib}.  Contrarily to the Young's modulus, the TCE recovers at $T_{ann}$=900 $^{\circ}$C. As the porous layers are not recovered at these annealing temperatures, it suggests that the non-porous layers are the main origin of TCE reduction.

From T$_{ann}$ = 700 $^\circ$C, the amplitude of $\Delta$TCE reduces significantly. At $T_{ann}$ = 900 $^{\circ}$C $\Delta$TCE is approximately -25\%. This could either mean that the recovered non-porous layers have not a similar molecular arrangement as the pristine material, resulting in a negative $\Delta$TCE. Another explanation is that the porous layers possess a positive $\Delta$TCE, since the porous layers still exists for these annealing conditions. Effectively this would mean that the TCE reduction of the non-porous layers is higher than what we effectively measured, hence TCE$_{nl}>$TCE$_{l}$.

Raman spectroscopy studies by Wang \textit{et al.}, Witcher \textit{et al.} and Zhang \textit{et al.} monitored the D$_{1}$ and D$_{2}$ peaks as function of the annealing temperature, these peaks are usually attributed to a three- and four-members rings structures \cite{Witcher2013-fl, Zhang2014-kt,Wang2021-ib}. Typically, they observed  $T_{relax}$ of the D$_{1}$ and D$_{2}$ peaks is in the range from 600 to 700 $^{\circ}$C. The annealing behavior of these densification signatures in the Raman spectrum correlates well with the annealing behavior of the TCE found in this work. In a previous work, based on a dedicated study related to the athermal Young's modulus, we suggested that the non-porous layers in the nanograting modification are densified \cite{Vlugter2020-xl}. This fact was also reported in earlier work by Bricchi \textit{et al.}, but this time, based on  birefringence measurements \cite{Bricchi2006-qy}. 

\onecolumngrid\
\twocolumngrid\

%These observations are also consistent with indirect observations reported by E. Bricchi \textit{et al.} \cite{Bricchi2006-qy}, related to the thermal stability of the nanograting's birefringent behavior- where the birefrigent behavior mostly is governed by the porous planes.

%For the non-porous layers, a lower energy activation energy is sufficient to  the pristine configuration, whereas for the nature of the porous structure indicates that higher activation energies are needed to recover the disrupted structure.

Let us now consider the annealing behavior for specimens densified with other non laser-based methods, namely, cold- \cite{Grimsditch1986-si} and hot-mechanical compression \cite{Guerette2018-fs, Cornet2019-cr}, fast neutron radiation \cite{Primak1958-pv} or shockwave induced densification \cite{Okuno1999-rh}. Hot-compressed and shockwave densification reveal a similar trends for densification versus $T_{ann}$, a plateau for the lower temperatures and a decrease at higher temperatures. However, for the cold-compressed methods (where SiO$_2$ is compressed beyond its 'plastic' limit of 9 GPa at room temperature \cite{Grimsditch1986-si}) and for fast-neutron radiation scenario, a monotonic decrease is revealed \cite{Primak1958-pv}. Shockwave-induced densification reveals a relaxation temperature at T$_{relax}\sim600^\circ$C \cite{Okuno1999-rh}, while for hot-compressed specimens, it depends on the magnitude of the applied compression pressure \cite{Guerette2018-fs} and temperature \cite{Cornet2019-cr}. Cornet \textit{et al.} studied the relaxation of hot-densified SiO$_2$ using small- and wide- X-ray diffraction methods \cite{Cornet2019-cr}. They revealed that for higher compression temperatures, the densified structure is more thermally stable (i.e. a higher T$_{relax}$). In their case, T$_{relax}$ was ranging between 400 $^\circ$C to 850 $^\circ$C for compression temperatures ranging between 400 $^\circ$C to 1020 $^\circ$C.

\begin{figure}[h!]
\centering
\includegraphics[width=1\linewidth]{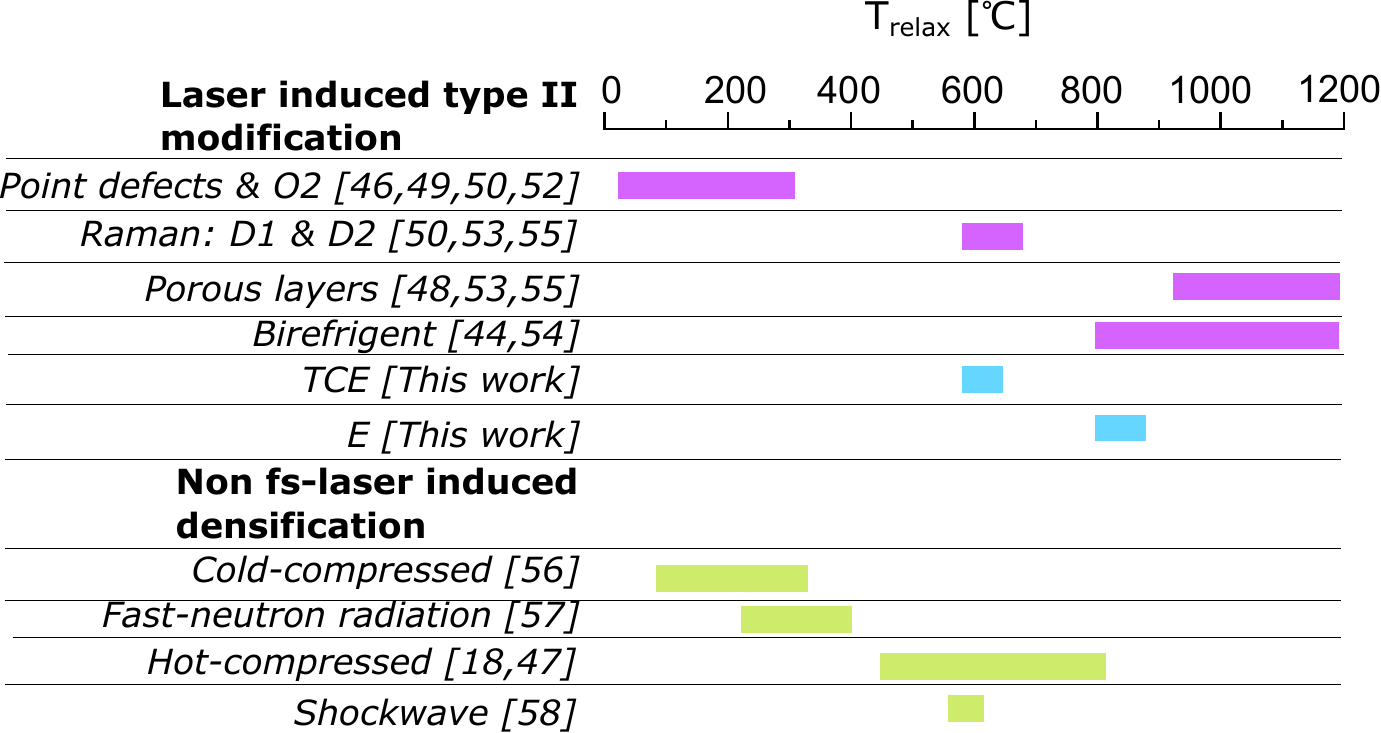}
\caption{An overview of the relaxation temperatures during annealing for laser induced material changes and for non-laser induced densifications in a-SiO$_2$. Great care should be taken when extracting precise values from this graph, as the exact values depend on the exact material composition (e.g. OH or Cl content) and annealing conditions (e.g. duration and heat-up/cool-down rate). Nevertheless, it provides a useful global overview.}
\label{Fig_SummaryAnnealingResults}
\end{figure}

The annealing behavior in this work correlates well with the annealing behavior of the densification signatures found in the Raman spectrum, as well with other non laser-based densification methods as can be seen in Figure \ref{Fig_SummaryAnnealingResults} that provides a visual summary. Summarizing these observations, the evidences are strong that the TCE reduction measured in this work is due to permanent densification where parts of the material is transformed to HDA. The latter is also mentioned in \cite{Wang2021-ib} . 

To summarize the preceding subsections, we have provided evidences that the non-porous layer is the main origin for the reduction of the TCE and not the porous layer. In this scenario, the elastically-deformed material due to volume changes do not play a role in the reduction of the TCE, neither do the point defects. We gave further evidences that the TCE reduction correlates well with permanent densification signatures, and therefore, we suggest that the nanograting modification includes volumes that are in a high density amorphous state. Additionally, we highlight that possible densification around nano-pores in the porous layer do not contribute to a measurable reduction of TCE as observed in this work. 

\section*{Discussion on the nature of the TCE reduction}
\begin{figure}[H]
\centering
\includegraphics[width=\linewidth]{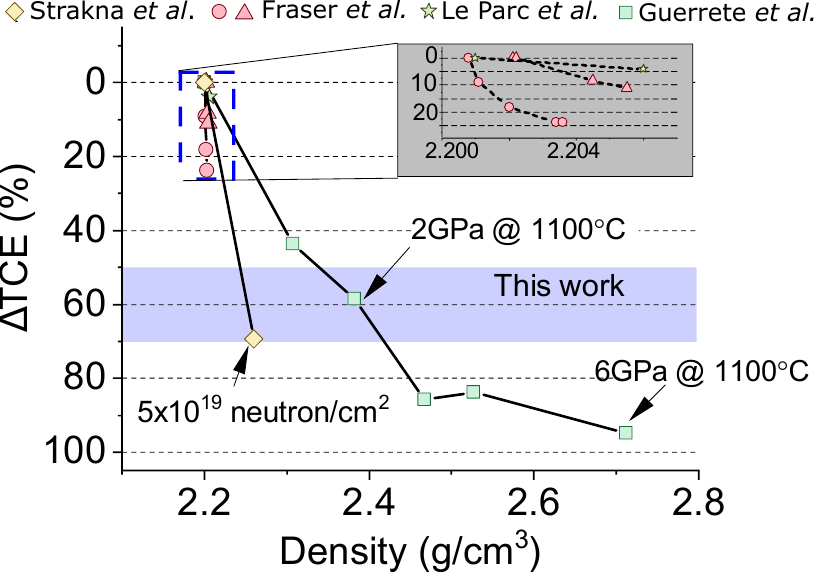}
\caption{A summary of $\Delta$TCE obtained by different densification process as reported in literature (Guerrete \textit{et al.} 2018 \cite{Guerette2018-fs}, Le Parc et al. 2006 \cite{Le_Parc2006-oz}, Fraser \textit{et al.} 1968 \cite{Fraser1968-xi}, Strakna \textit{et al.} 1963, \cite{Strakna1963-tu}). The typical value obtained in this study is also shown.}
\label{Fig_LitReviewTCE}
\end{figure}

\noindent Various works have demonstrated that densifying SiO$_2$ reduces the TCE. Le Parc obtained a TCE reduction of thermally quenching a-SiO$_2$  at atmospheric pressure \cite{Le_Parc2006-oz}. Based on Brillouin scattering measurement, the author estimated a TCE reduction of 5\% maximum. Strakna \textit{et al.} studied the reduction of TCE for fast-neutron irradiated a-SiO$_2$ samples in the temperature range from 100K to 300K \cite{Strakna1963-tu}. They showed that the TCE is reduced by approximately 70\%. In recent work, Guerrete \textit{et al.} investigated the TCE for hot-compressed a-SiO$_2$ \cite{Guerette2018-fs} and found a TCE reduction of more than 80\%, for compression-pressures of 6 GPa at a temperature of 1100 $^\circ$C. Figure \ref{Fig_LitReviewTCE} summarises the reduction of TCE obtained with various densification processes, compared with the result of this study (in the particular case of type II modification, a deposited energy of 160 J/mm$^2$ and a polarisation perpendicular to the laser-scanning direction).  

%For ionizing type of radiations, a more gentle reduction is found, approximately 6\%, like in the case of $\gamma$ radiation.
Based on this overview, the evidences are strong that compacting a-SiO$_2$ results in a reduction of the TCE. In a densified material, the compaction hinders conformation changes of the six-membered rings structures, due to the lack of space \cite{Guerette2018-fs,Huang2017-lk}. Furthermore, the ring size distribution seems to have an effect on the magnitude of the TCE reduction. For instance, densification induced by ionizing radiation,  three and four ring-members are generated at the expense of a decay of the number of six-membered ring structures \cite{Vukelic2010-mr,Shcheblanov2016-st}. As  the  three- and four- membered rings are stiffer structures, they are less prone to conformation changes \cite{Huang2017-lk} and consequently, no stiffening effect upon a temperature increase is expected for lower number ring-members, as this would be the case for six-membered rings in fused silica. Works on intermediate glass\cite{Rabia2016-zf,Huang2017-lk} suggest a correlation between the magnitude of the TCE and the amount of the six-membered rings \cite{Rabia2016-zf,Huang2017-lk}.

The TCE reduction obtained in this study seems remarkably high, when comparing it with results obtained for the homogeneous modification (Type I), Figure \ref{Fig_Mainresults}a. This indirectly suggests that the densification of the non-porous planes in the nanograting modification is higher than homogeneous modification. This is in other works also indirectly suggested, based on the study to the athermal elastic properties \cite{Vlugter2020-xl} and refractive index \cite{Bricchi2006-qy} of the non-porous planes. Where both the athermal elastic properties and refractive index is significantly higher compared to the one of the Type I modification. 

The onset for densification of the Type I modification is generally attributed to fast cooling after laser exposure \cite{Wang2021-ib} or/and due to a bond-breaking mechanism, induced by ionization radiation, creating smaller ring-members  \cite{Shcheblanov2016-st}. As reviewed above, both mechanisms, fast thermal quenching and ionizing radiation, produce similar TCE reductions as obtained for the homeogenous modification. The result correlates well with similar densification mechanisms. Simultaneously, this raises the question what other mechanism is taking part in reducing the TCE of the nanograting regime, and this with a magnitude of approximately 60\%. 
In \cite{Vlugter2020-xl}, we envisioned that the nanograting modification's non-porous planes densification mechanism is different to the one found in type I exposure conditions. Accordingly, we further suspected that pressures generated at the porous-plane location further densify the surrounding material, i.e. the non-porous planes, at elevated temperatures as the volume change of the  porous planes exert a static pressure. Based on the new insights provided here, these hypotheses might not hold. From the annealing experiments, in this work, we can state that this static pressure of the porous planes does not play a role. The TCE reduction, and therefore the densification of these zones, recovers for $T_{ann}>$600 $^\circ$C, while the static pressure of the porous planes is still present. This observation suggests that the onset of the densification is of temporal nature and appears during the exposure or shortly after.

A precise justification for the onset for higher densification in the non-porous planes is beyond the scope of this work and deserves a more dedicated study. We would like nevertheless to point-out a possible line of thoughts. The nano-pores found in the porous layers are a product of glass decomposition due to the locally enhanced plasma density \cite{Bhardwaj2006-so,Lancry2013-fv}. The localized enhanced electron density \cite{Rudenko2016-ob} leads to a rapid, sub-nanosecond \cite{Bulgakova2015-um} increase of the lattice temperature. A shockwave is generated due to fast and local thermal expansion of the material. One one hand, the negative pressure caused by the rarefaction wave, results in a tensile force onto the material and initiates a cavitation process, leading to the formation of nanopores \cite{Rudenko2018-iq}.  On the other hand, at the front of the shockwave, compressive stress exists. This compressive stress might be the cause of this high TCE reduction found in non-porous layer.   

\section*{Conclusion}
\noindent With our earlier works, \cite{Vlugter2019-ye} that described the behavior of the coefficient of thermal expansion and \cite{Vlugter2020-xl} that addressed the change in Young's modulus at room temperature, this work finalizes a thermomechanical description of femtosecond irradiation effects on a-SiO$_2$, by providing insights related to the laser-induced changes induced to the temperature coefficient of elasticity (TCE). 

Here, we observed that the local decrease of the temperature coefficient of elasticity of a-SiO$_2$ exposed to femtosecond laser irradiation can be significant, eventually reaching 60\% and among the highest reported effects for the various methods reported to decrease the TCE of silica.

These findings demonstrate the opportunity of tailoring the thermomechanical behavior of dielectrics in a non-contact and localised manner. This is useful for e.g. watchmaking,  frequency references and precision devices which requires high dimensional stability. Furthermore, the applied methodology in this work can be used for exploring the TCE reductions in other glass formers using ultrafast laser.

\section*{Acknowledgments}
\noindent The Galatea Lab acknowledges the sponsorship of Richemont International SA.

\appendix
\numberwithin{equation}{section}
\numberwithin{figure}{section}
\section{Derivation of resonance frequency - material property relation}

\begin{figure}[H]
\centering
\includegraphics[width=\linewidth]{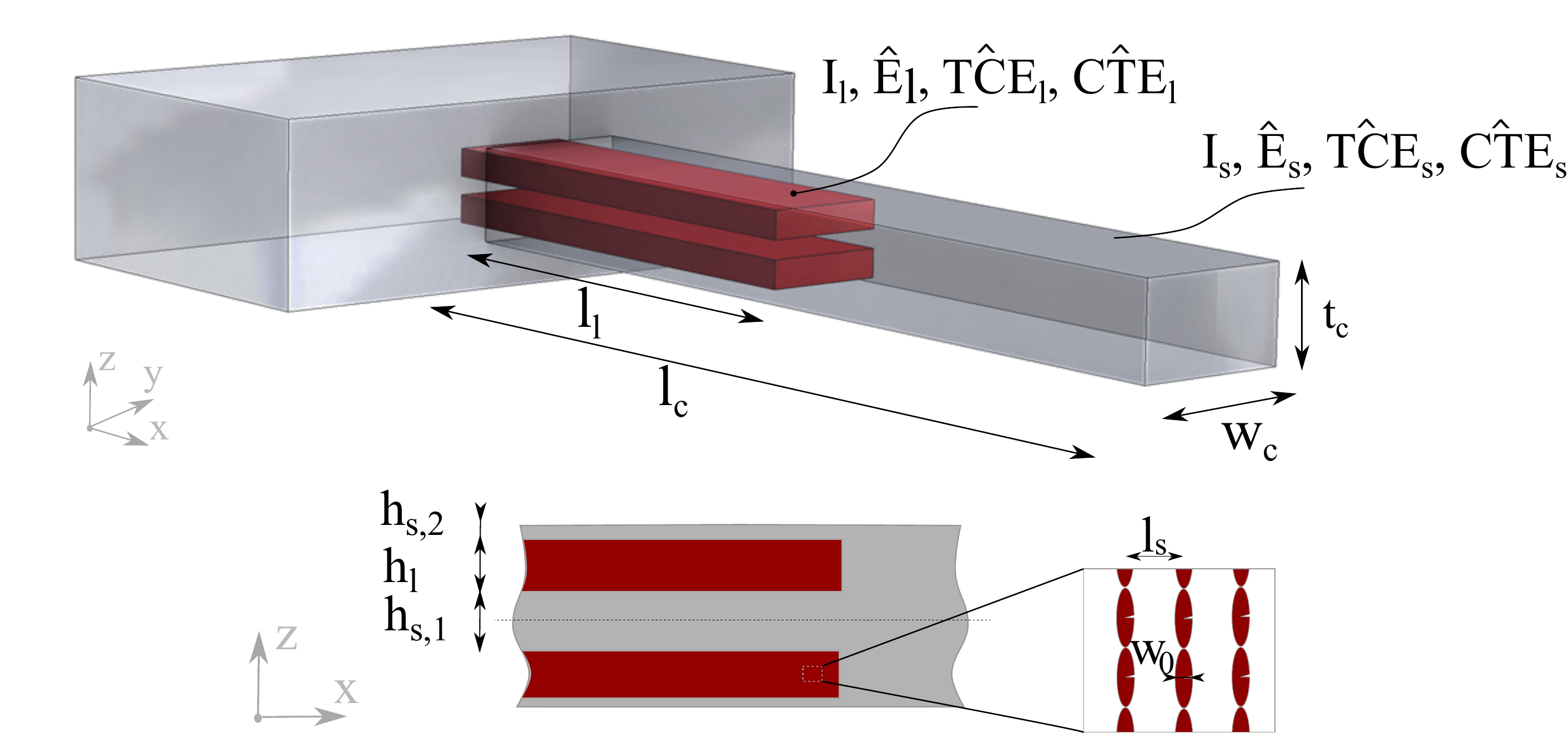}
\caption{A schematic drawing of the resonator with the LAZ in red. The dimensions, second moment of area $I_{\cdot}$ and material properties ($E_{\cdot}$, TCE$_{\cdot}$, CTE$_{\cdot}$). }
\label{ResonatorSchematic}
\end{figure}
\noindent In this Appendix we express the resonator's resonance frequency in terms of material properties of the laser exposed and pristine material, since it is exposed partly, see Figure \ref{ResonatorSchematic}.
The lowest resonant frequency of a clamped-free cantilever is expressed in equation \ref{equa_frequency1}. $\rho$ and $V$ are the mass density and the volume of the cantilever. E. and I. are the Young's modulus and the second moment of inertia, respectively, with the subscripts .$_l$ and .$_s$ referring to the laser affected zones (LAZ) and the non-exposed ones, respectively.  The circumflex as diacritical mark indicates that the variable is a volume average and the subscript .$_0$ indicates a temperature invariant variable. $\lambda$ is the wavenumber, which corresponds to a resonance mode.  

\begin{equation}
    f=\frac{\lambda^2}{2\pi}\sqrt{\frac{\hat{E}_{l}I_{l}+E_{s}I_{s}}{\rho V l^3}}
    \label{equa_frequency1}
\end{equation}

with $\rho=\rho_0/(1+\textrm{CTE}\Delta T)$, $V=V_0(1+\textrm{CTE}\Delta T)$, $l=l_0(1+\hat{\textrm{CTE}}_l\Delta T)$, $I_{l}=I_{l,0}(1+\hat{\textrm{CTE}}_l\Delta T)^4$ and $I_{s}=I_{s,0}(1+\textrm{CTE}_s\Delta T)^4$, equation \ref{equa_frequency2} becomes

\begin{equation}
    f=\frac{\lambda^2}{2\pi}\sqrt{\frac{\hat{E}_{l} I_{l,0} (1+\hat{\textrm{CTE}_l}\Delta T) + E_{s} I_{s,0} (1+\textrm{CTE}_{s}\Delta T)}{\rho_0 V_0 l_0^3}}
    \label{equa_frequency2}
\end{equation}

The temperature-dependent elastic constants are linearized to the first order, i.e. $E_{s}=E_{s,0}+dE_{s}/dT \Delta T$. The thermal derivative of the frequency, evaluated at $\Delta T=0$, is;

\begin{eqnarray}
\frac{\partial f}{\partial T} \bigg\rvert_{\Delta T=0}=\frac{\lambda}{4\pi}\frac{1}{\sqrt{\hat{E}_{l,0}I_{l,0}+E_{s,0}I_{s,0}}}\frac{1}{\sqrt{l_0^3\rho_0V_0}}  \cdots \nonumber\\
\left[I_{l,0}\left(\frac{d\hat{E}_{l}}{dT}+\hat{E}_{l,0}\hat{\textrm{CTE}}_l\right)+I_{s,0}\left(\frac{d\hat{E}_{l}}{d T}+E_{s,0}\textrm{CTE}_s\right)\right]
\end{eqnarray}

Simplifying further the equation, with the assumption that the terms containing the coefficient of thermal expansion ($\textrm{CTE}_{.}$) are much smaller (by an order of magnitude $\sim 4$) than with the ones containing elastic constant thermal derivatives ($\partial E_\cdot/\partial T$), gives

\begin{equation}
\frac{\partial f}{\partial T} \bigg\rvert_{\Delta T=0}=\frac{f_0}{2}\frac{1}{\hat{E}_{l,0}I_{l,0}+E_{s,0}I_{s,0}}\left[I_{l,0}\left(\frac{d\hat{E}_{l}}{d T}\right)+I_{s,0}\left(\frac{d E_{s}}{d T}\right)\right]
\label{equa_TCF_withoutCTE}
\end{equation}

rewriting the equation in terms of $\textrm{TCE}_{.}$, i.e. $dE_{l}/dT=E_{l,0}\textrm{TCE}_l$, gives,

\begin{equation}
\frac{\partial f}{\partial T} \bigg\rvert_{\Delta T=0}=\frac{f_0}{2}\left[\frac{I_{l,0}\hat{E}_{l,0}\hat{\textrm{TCE}}_l+I_{s,0}E_{s,0}\textrm{TCE}_s}{\hat{E}_{l,0}I_{l,0}+E_{s,0}I_{s,0}}\right]
\label{equa_TCF2}
\end{equation}

$\hat{\textrm{TCE}}_l$ and $\hat{E}_{l,0}$ are the average properties of the modified volume, which are further specified in equations \ref{Equa_beta} and \ref{Equa_Ehat}, in terms of the volume fractions $V_l$ and $V_s$. Where LAZ and pristine volume fraction is expressed as $V_l=w_0/l_s$ and  $V_s=1-V_l$ (see Figure \ref{ResonatorSchematic}).

\begin{equation}
\hat{\textrm{TCE}}_l     =      \frac{\textrm{TCE}_s V_sE_{l,0}+\textrm{TCE}_l V_lE_{s,0}}{V_sE_{l,0}+V_lE_{s,0}} \label{Equa_beta}
\end{equation}

\begin{equation}
\hat{E}_{l,0}         =      \frac{E_{l,0}E_{s,0}}{E_{l,0}V_s+E_{s,0}V_l} \label{Equa_Ehat}
\end{equation}

\section{Derivation of bimorph stress relation}

\begin{figure}[H]
\centering
\includegraphics[width=\linewidth]{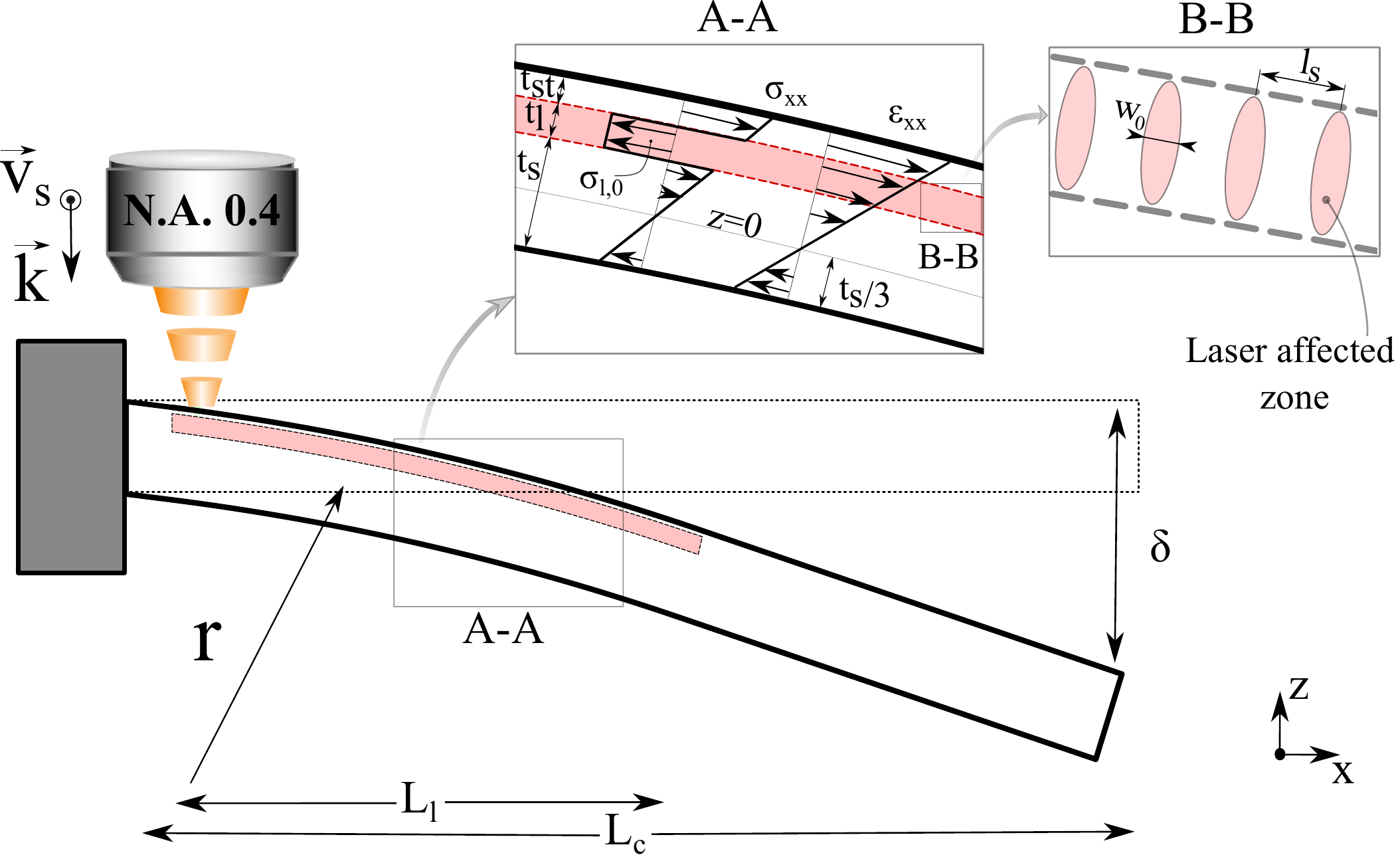}
\caption{A schematic of a laser-exposed  cantilever, with the dimensions indicated as well as the stress and strain profiles across the thickness. $\hat{\sigma}_{l,0}$  is the stress in laser exposed layer.}
\label{Fig_LaserexposedCantilever}
\end{figure}

 \noindent In this annex, we derive a mathematical expression for the stress in the modified layer $\hat{\sigma}_{l,0}$ of the bimorph structure. In Figure \ref{Fig_LaserexposedCantilever} a schematic is depicted of a cantilever being exposed just below the top-layer and consequently bend with a bending radius $r$ due to volume expansion of the laser affected zone. The stress and strain profile is illustrate in the inset A-A Figure \ref{Fig_LaserexposedCantilever}. Where the stress in the laser affected layer, $\hat{\sigma}_{l,0}$, is opposite in sign compared with the stress in the pristine bulk. For this system, the stress and strain in the pristine bulk is zero at a distance $t_s$/3 from the bottom surface, defining the neutral-axis \cite{Hsueh2002-qj}. We define the neutral-axis as the origin of the z-coordinate, i.e. z=0.      
 
 The relationship between bending radius and strain is;
 
  \begin{equation}
\epsilon_{xx}(z)=\frac{z}{r}
\end{equation}
 
 Newton's third law is used to analytical determine $\hat{\sigma}_{l,0}$, see equation \ref{Equa_ImprovedStoneysEquation}. The first two terms represents the force in pristine bulk, with $E_s$ as the Young's modulus of this volume. Where the last term represent the force in the LAZ.  
 
 \begin{equation}
0=\frac{\partial \epsilon_{xx}(z)}{\partial z}E_s\left(\int_{-t_s/3}^{t_s2/3} z\cdot\textrm{d}z+\int_{t_s2/3+t_l}^{t_s2/3+t_l+t_{s,t}} z\cdot\textrm{d}z\right)+\hat{\sigma}_{l,0}t_l
\label{Equa_ImprovedStoneysEquation}
\end{equation}
Rewriting equation \ref{Equa_ImprovedStoneysEquation} gives the expression for $\hat{\sigma}_{l,0}$;  

\begin{equation}
\hat{\sigma}_{l,0}=-\frac{E_st^2_s}{6t_lr}-\frac{E_s}{t_lr_0}\left[\frac{1}{2}\left(t_s\frac{2}{3}+t_l+t_{s,t}\right)^2-\frac{1}{2}\left(\frac{t_s}{3}+t_l\right)^2\right]
\label{Equa_AVGSIGMA_L}
\end{equation}

Note, that the first term is the Stoney's equation.

%\bibliography{bibliography}

\end{document}